\documentclass[twocolumn]{aastex62}

\newcommand\pf{PKS~1510$-$089}
\newcommand\pfs{PKS~1510$-$089\textrm{ }}

\graphicspath{{./}{figures/}}

\received{August 7, 2019}
\revised{August 19, 2019}
\accepted{August 22, 2019}
\submitjournal{ApJL}

\shorttitle{\emph{Fermi} detection of IC/CMB in Large-scale Jets}
\shortauthors{Meyer et al.}

\begin{document}

\title{The Origin of the X-ray Emission in Two Well-Aligned Extragalactic Jets: The Case for IC/CMB}

\correspondingauthor{Eileen T. Meyer}
\email{meyer@umbc.edu, eileen.meyer@gmail.com}

\author[0000-0002-7676-9962]{Eileen T. Meyer}
\affil{Department of Physics, University of Maryland, Baltimore County, 1000 Hilltop Circle, Baltimore, MD 21250, USA}

\author{Adurshsiva R. Iyer}
\affil{Department of Physics, University of Maryland, Baltimore County, 1000 Hilltop Circle, Baltimore, MD 21250, USA}

\author[0000-0001-9018-9553]{Karthik Reddy}
\affil{Department of Physics, University of Maryland, Baltimore County, 1000 Hilltop Circle, Baltimore, MD 21250, USA}

\author{Markos Georganopoulos}
\affil{Department of Physics, University of Maryland, Baltimore County, 1000 Hilltop Circle, Baltimore, MD 21250, USA}
\affil{NASA Goddard Space Flight Center, 8800 Greenbelt Road, Greenbelt, MD 20771, USA}

\author{Peter Breiding}
\affil{Department of Physics \& Astronomy, West Virginia University, Morgantown, WV 26506, USA}

\author{Mary Keenan}
\affil{Department of Physics, University of Maryland, Baltimore County, 1000 Hilltop Circle, Baltimore, MD 21250, USA}



\begin{abstract}
Over the past two decades, the most commonly adopted explanation for high and hard X-ray emission in resolved quasar jets has been inverse Compton upscattering of the Cosmic Microwave Background (IC/CMB), which requires jets which remain highly relativistic on 10-1000 kpc scales. In more recent years various lines of observational evidence, including gamma-ray upper limits, have disfavored this explanation in favor of a synchrotron origin. While the IC/CMB model generally predicts a high level of gamma-ray emission, it has never been detected. Here we report the detection of a low-state \emph{Fermi/LAT} gamma-ray spectrum associated with two jetted AGN which is consistent with the predictions of the IC/CMB model for their X-ray emission. 
We have used archival multiwavelength observations to make precise predictions for the expected minimum flux in the GeV band, assuming that the X-ray emission from the kpc-scale jet is entirely due to the IC/CMB process. In both sources -- OJ 287 and PKS 1510-089 -- the minimum-detected gamma-ray flux level agrees with predictions. Both sources exhibit extreme superluminal proper motions relative to their jet power, which argues for the well-aligned jets required by the IC/CMB model.  In the case of PKS~1510-089, it cannot be ruled out that the minimum gamma-ray flux level is due to a low state of the variable core which only matches the IC/CMB prediction by chance. Continued long-term monitoring with the \emph{Fermi}/LAT could settle this issue by detecting a plateau signature in the recombined light-curve which would clearly signal the presence of a non-variable emission component.
\end{abstract}

\section{Introduction} \label{sec:intro}

In August of 1999, the newly-launched \emph{Chandra} X-ray observatory observed the distant ($z=0.651$) quasar PKS~0637$-$752, during the orbital checkout and activation phase of the mission \citep{chartas2000,schwartz2000}, unexpectedly discovering X-ray emission from the resolved kpc-scale jet. Over the intervening years, dozens more X-ray bright jets have been discovered in which, as in PKS~0637$-$752, the X-ray emission has a hard spectrum and high flux level which clearly indicates a second emission component from the radio-optical synchrotron emission. The nature of this X-ray emission, while at some points thought understood, has had no clear identification for nearly 20 years. The few possible mechanisms imply vastly different physical conditions in the jet and total energy budgets. 

Shortly after the original discovery of PKS~0637$-$752, \cite{tav00} and \cite{cel01} independently suggested that the X-ray emission could be explained by inverse-Compton upscattering of the Cosmic Microwave Background (IC/CMB). This explanation was originally considered and discarded in the discovery papers by \cite{chartas2000} and \cite{schwartz2000}; those authors assumed \citep[as is consistent with population observations generally, e.g.][]{arshakian2004} that the jet would not remain highly relativistic at 100 kpc or more from the central engine. Under only mildly relativistic conditions it is impossible to reproduce the high X-ray flux level with IC/CMB without unreasonable (many orders of magnitude) deviations from equipartition. However, \cite{tav00} found that by taking the large-scale jet to be very well aligned ($\theta<6^\circ$) and highly relativistic ($\Gamma>18$), the IC/CMB emission could be Doppler-boosted enough to match the X-ray observations, though at a cost of requiring a jet power 10 times the Eddington value. Many other jets with a second (hard) spectral component emerging in the X-rays have been modeled as cases of IC/CMB \citep[e.g.,][]{sambruna2002, sambruna2004, marshall2005, jorstad2006, tavecchio2007, marshall2011, perlman2011, kharb2012, stanley2015}. The vast majority of these jets were in sources identified as quasars, i.e., aligned counterparts of FR II radio galaxies, which are typically more powerful than FR I radio galaxies or BL Lacs (their well-aligned blazar counterparts). OJ~287 is a notable example as an FR~I \citep{marscher2011}. 

While IC/CMB has been the most popular explanation for the hard X-ray emission in resolved jets, it has never been positively confirmed as the correct one. A few years before the launch of \emph{Fermi}, in a clear-sighted review of the state of the case, \cite{hardcastle2006} noted the many problems for the IC/CMB model, including the inconsistency of the highly relativistic jet speeds on hundreds-of-kpc scales with population statistics, the problem of X-ray/radio knot offsets and fine-tuning of the required minimum electron Lorentz factor. As an alternative, he suggested that a second-synchrotron model (i.e., emission from a heterogeneous rather than `one zone' model) be reconsidered. A few other authors have also preferred a second synchrotron component for some cases \citep{dermer_atoyan,miller2006}.
The difficulty up to this point was that it was not possible to use spectral energy distribution (SED) modeling of radio to X-ray data to distinguish the two scenarios. Even apparent discrepancies between radio and X-ray spectral indices can be accommodated. The spectral indices, particularly in the X-rays, are rarely well-constrained. In addition they trace very different particle populations under the IC/CMB model, with the X-ray producing electrons having a lower particle Lorentz factor ($\gamma$) by a factor of 10 or more.

\cite{geo06} noted that the IC/CMB model for the X-ray emission implies a very high level of gamma-ray emission which generally peaks at or near the \emph{Fermi}/LAT (then GLAST) band, and that the level of this emission is also completely predicted by the radio-optical spectral shape and the level of the X-ray emission. With a well-sampled radio-optical synchrotron SED, there is no freedom about the level of gamma-ray emission \citep[see also][]{meyer17}. Since the launch of \emph{Fermi} in 2008, we have been looking for this steady gamma-ray emission from second-component X-ray jets. Up until now we have not found any evidence for it, ruling out the IC/CMB model in dozens of sources (\citealt{meyer14,meyer15,meyer17,breiding2017,breiding2018_thesis}; Breiding et al., 2019, in prep.).
 
 In this paper we report the first detection of a low-state gamma-ray spectrum with the Fermi/LAT which is consistent with the IC/CMB predictions based on the assumption that the X-ray emission is entirely due to IC/CMB. The two sources are considerably different from one another in terms of jet power; OJ~287 is a low-power BL Lac object while \pfs is a powerful quasar.  After presenting the observations we will discuss the likely reasons that these jets are atypical compared to the majority of X-ray jets where the X-rays are not dominated by IC/CMB.  

In this paper we assume the current standard $\Lambda$CDM cosmology with Hubble constant $H_0$=73~km~s$^{-1}$~Mpc$^{-1}$, $\Omega_M$= 0.27, and $\Omega_\Lambda$=0.73. At the distance of OJ~287 ($z$ = 0.306) 1$''$ corresponds to a projected distance of 4.36 kpc, while for \pfs ($z=0.36$) it corresponds to 4.87 kpc.

\begin{figure*}[t]
    \centering
    \includegraphics[width=6.5in]{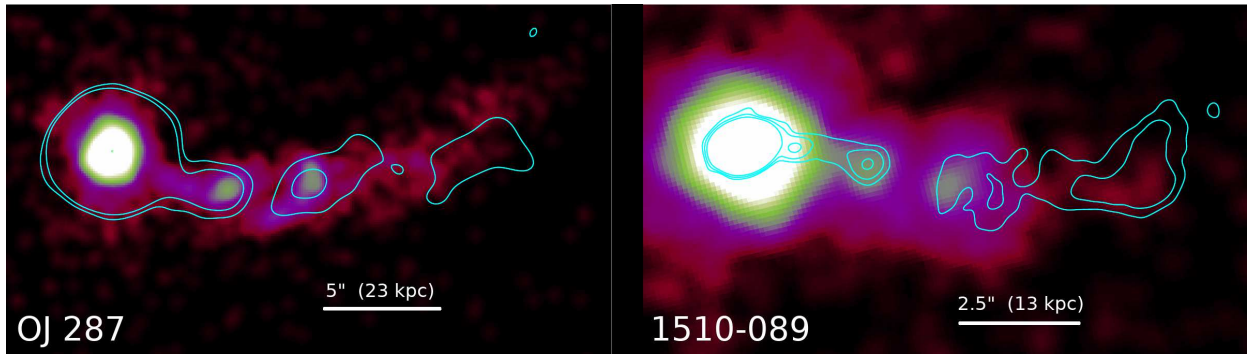}
    \caption{Chandra X-ray imaging of OJ~287 and \pfs with radio contours overlaid. OJ 287 is in North-up orientation while \pfs has been rotated for comparison. The (0.4-8 keV) \emph{Chandra}/ACIS observations are from Observation IDs 9182 and 11432; they have been reprocessed using CIAO ver. 4.11 and have been rebinned to 1/5 native scale. The radio contours shown on the OJ~287 image are from L-band A-configuration VLA observation AC108 (clean beam size 1.69$''$ $\times$ 1.59$''$)  while \pfs are from C-band A-configuration VLA observation AO070 (clean beam 0.51$''$ $\times$ 0.39$''$).     }
    \label{fig:X-ray}
\end{figure*}

\section{Data Sources and Methods}
\subsection{VLA and ALMA}

For OJ~287, the radio fluxes for the entire large-scale jet at 1.4 GHz, 4.8 GHz, and 8.6 GHz were taken from \cite{marscher2011}. This source has also been utilized as a calibrator source for ALMA observations; we analyzed relatively deep Band 3 observations taken from project 2016.1.00406.S where OJ 287 served as phase calibrator, but the jet was not detected -- likely because of dynamic range limitations given the very bright core (5.5 Jy at 100 GHz). 

For \pf, we analyzed historical VLA observations at L and C band in A-configuration from project codes AH938 and AO070, respectively. The data were calibrated using standard procedures in CASA version 5.3.0; as the source is very bright \pf~was used as its own phase calibrator. Several rounds of phase-only (non-cumulative) and a final amplitude and phase self-calibration were applied to the data after successive deconvolutions with CLEAN as is standard practice for bright compact sources. The final RMS of these images were 4.5 $\times$ 10$^{-4}$~Jy and 1.9 $\times$ 10$^{-4}$~Jy, with synthesized beam sizes of 1.56$''$ $\times$ 1.15$''$ and 0.51$''$ $\times$ 0.39$''$, respectively.

\pfs has also been observed by ALMA; we reduced archival band 4, 6, and 7 observations from project 2016.1.00116.S. The data were initially processed using the `scriptForPI.py' script with CASA in pipeline mode. In all cases, after initial imaging we applied 1-2 rounds of phase-only self-calibration before a final round of amplitude and phase self-calibration. The final RMS of these images was 7.3 $\times$ 10$^{-5}$~Jy, 1.0 $\times$ 10$^{-4}$~Jy and 1.6 $\times$ 10$^{-4}$~Jy with synthesized beam sizes of  2.49$''$ $\times$ 1.72$''$, 1.82$''$ $\times$ 1.36$''$, and 1.20$''$ $\times$ 0.93$''$, respectively.

For all radio imaging analyzed here we produced `core-subtracted' images in order to better isolate the total flux from the extended jets. To do so we used CLEAN to populate the `model' column of the post-self-calibration CASA MS file with components only at the location of the core and then subtracted these from the visibility data using the CASA task \texttt{uvsub}. We then ran a final round of CLEAN on the now-subtracted MS file to produce the image without core emission.

\subsection{HST and Chandra}
Both jets have been observed in the optical by HST, with only upper limits to the jet emission as reported by previous authors \citep{marscher2011,sambruna2004}, which we have included in our SEDs.

In the X-rays, OJ~287 has been observed once by \emph{Chandra} in 2007 December and the flux of the entire jet is reported in \cite{marscher2011} as 3.06 $\pm$ 0.09 $\times$ 10$^{-2}$ cts s$^{-1}$ from 0.2$-$6~keV. Using their measured spectral index for the entire jet of $\alpha$ = 0.61 $\pm$ 0.06 we have converted this to 10.4 $\pm$ 0.3~nJy at 1~keV.

\emph{Chandra} observations of \pfs were first reported by \cite{sambruna2004} based on a 9~ks observation from 2001 March. We re-analyzed a deeper 45~ks ACIS-S observation from (OBSID: 11432), which was taken on 2010 April 5 using the FAINT telemetry mode. The analysis was conducted using CIAO 4.11. After standard reprocessing using CALDB 4.8.2, we screened the data for background flares (there were none) and extracted the data from 0.4-8 keV with a final exposure time of 43.38~ks. 

\begin{figure*}[t]
    \centering
    \includegraphics[width=7in]{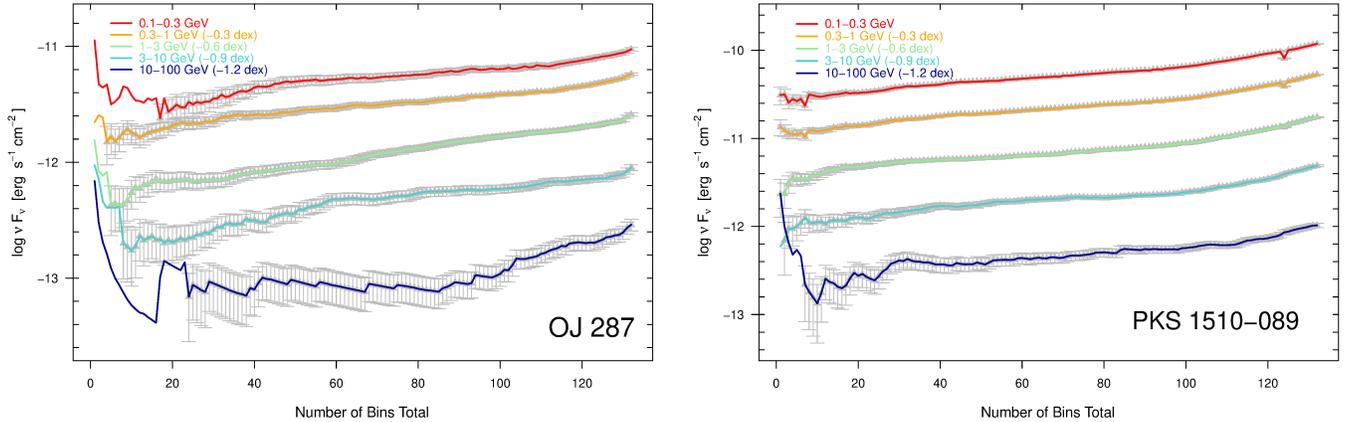}
    \caption{Results of the progressive binning \emph{Fermi}/LAT analysis for OJ~287 (left) and \pf, (right). We have plotted the $\nu F_{\nu}$ flux versus total number of bins combined separately for each energy bin. The curves have been multiplied by factors as shown in the legend in order to space them out for clarity (the lowest energy bin is at the actual flux level observed). In both cases 95\% upper limits are indicated by a lack of error bars while detected fluxes (where the band TS has reached at least 10) are shown with gray error bars. }
    \label{fig:Light Curves}
\end{figure*}

The first radio knot in the jet of \pfs is located 1.3$''$ from the core location, which means that the inner jet is significantly affected by the wings of the PSF from the bright X-ray core. We conducted PSF simulations of the core in order to account for its contribution of flux to the inner jet. We used the toolset MARX, which provides ray tracing routines for the Chandra optics to simulate the PSF. We provided the aspect solution file from the observation and the spectrum extracted using \textit{specextract} from a 1.5'' (95\% ecf) radius centered on the core as inputs to MARX via the \textit{simulate\textunderscore psf} module in CIAO. The spectrum was modeled using an absorbed power-law model in SHERPA with the absorption column density set to a sum of an intrinsic column density and a Galactic column density in the direction of the quasar, the latter set fixed at $n_H$ = 7.13 $\times$ 10$^{20}$ cm$^{-2}$. The nuclear spectrum was well-fit by this model (reduced $\chi^2$ = 0.85) with a power law index $\Gamma$ = 1.75 $\pm$ 0.01. After performing 100 iterations of \textit{simulate\textunderscore psf}, a final model of the PSF was obtained by merging the outputs. The event file thus obtained was then binned to 0.2 times the native ACIS-S resolution in the 0.4-8~keV range to obtain a flux image at an effective energy of 1~keV. We then subtracted the flux from the inner knot of the simulated psf from the total flux of the jet resulting in an estimate of 14.8 $\pm$ 0.2~nJy at 1~keV. This is slightly less than the reported sum of knots in \cite{sambruna2004} of 19.6 $\pm$ 1.9~nJy. This is likely due to differences in measuring the total flux (we used a single contiguous region rather than 1$''$ radius circular regions on individual knots) and method of accounting for core contamination. We also fit the extracted jet spectrum as a power-law in SHERPA, yielding a spectral index $\alpha$ = 0.56 $\pm$ 0.11, which agrees with the \cite{sambruna2004} values for the individual knots within their (rather large) errors. Sub-pixel (1/5) binned \emph{Chandra} images of the jets are shown in Figure~1 with radio contours overlaid. 

\subsection{\emph{Fermi/LAT}}

We utilized the progressive binning method to search for the minimum flux or upper limit in the five canonical \emph{Fermi}/LAT energy bands (0.1-0.3 GeV, 0.3-1 GeV, 1-3 GeV, 3-10 GeV and 10-100 GeV). The method used is identical to that described in \cite{meyer14} and \cite{meyer15} which we refer to for further details. Briefly, we used a 7 degree region of interest (ROI) to isolate photons observed by the \emph{Fermi}/LAT from the direction of the source over the time available at the time of analysis. For OJ 287 the mission elapsed time (MET) start and stop times are 239557417 to 577782027 and for \pfs they are 239557417 to 551014632. Using standard Fermi tools (version v11r5p3-fssc-20180124 for \pf, and conda-distributed version 1.0.2 for OJ~287) and the latest instrument response (P8R3\_V2) a light-curve was made for each source using bins of 1 week total Good Time Interval (GTI) time and an energy range of 100 MeV to 100 GeV. The light-curve was then reordered by test statistic (TS) value from lowest to highest value (where the TS is roughly significance squared). We then conducted a standard Fermi likelihood analysis on the source position for progressively combined bins -- first the lowest two, the lowest three, etc until all bins were analyzed together (the latter giving the average spectrum over the 10.5 years of Fermi observations). At each step in the recombined binning the flux (or upper limit) in each of the five \emph{Fermi} energy bands was calculated.  We then adopt the absolute minimum flux or upper limit in each band individually over the whole recombined binning to generate the minimum SED. In addition, for illustrative purposes we measured a `high state' composite spectrum by taking the ten highest bins (by TS value) in combination and producing the five band fluxes via maximum likelihood.

\section{Results}
The results of the \emph{Fermi}/LAT progressive binning analysis are shown in Figure~2. Here we plot either the 95\% upper limit (no error bars) or detected flux level (with error bars) for the source in each of the five energy bands versus the total number of bins combined (here the curves have been multiplied by regular factors as noted in the legend to space them out in the figure). As described above, the bins from the standard light-curve are first re-ordered from lowest to highest TS before being recombined sequentially as the two lowest, three lowest, etc for what we call the progressive binning analysis. Because of the low angular resolution of the \emph{Fermi}/LAT ($\gtrsim$~1$^\circ$) compared to the jets (tens of arcseconds), the fluxes and upper limits derived here apply to the core and large-scale jet combined.

\begin{figure*}[t]
    \centering
    \includegraphics[width=7in]{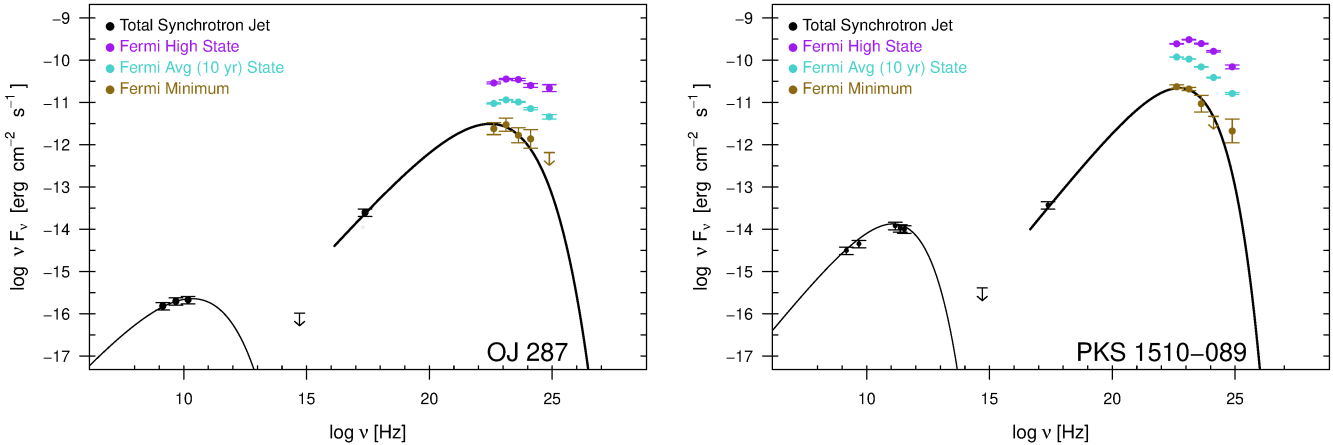}
    \caption{The SEDs for OJ~287 (left) and \pfs (right). The black data points (including X-rays) are the total flux in the large-scale jet, fit with a phenomenological model as in \cite{breiding2017}. The high-energy spectrum is the shifted IC/CMB curve as constrained by the synchrotron fit and the requirement to match the X-ray flux level. In the gamma-rays, three states of the source are shown as seen by the Fermi/LAT. The high state was measured by combining the highest ten (1-week) bins in the original light-curves. The `average' state is simply the total ten-year data set, while the `minimum' state is that derived from the progressive binning method described in the text. }
    \label{fig:SEDs}
\end{figure*}

The behavior observed in Figure~2 is slightly different for OJ~287 and \pf. For OJ~287, the source is initially not detected in any band. As the time-on-source increases, the flux limits generally decrease as expected (this is most dramatic in the highest energy band where the background is essentially zero) before the source is detected. For a source completely dominated by the variable core, the expected course after detection is to see steadily increasing fluxes since we have ordered the bins on significance (TS). However when dominated by a steady source, the expectation is that the detected flux level will remain steady while the error bars shrink. In the case of OJ~287, we do see signs of such `plateau' behavior consistent with a steady flux level in the highest energy band, with hints of a much smaller plateau in the middle three energy bands. This behavior is consistent with reaching a `floor' in the flux level due to the IC/CMB emission which is expected to be completely non-varying. 

For \pf, the source is detected in every band except the last two even from the shortest combined bin, as shown at right in Figure~2. While the flux in each band is somewhat slow to rise, there is no clear sign of a plateau in any band. Thus either we are just barely able to detect the IC/CMB flux level before being overtaken by the brighter core, or the minimum-flux state of the core is coincidentally at the level expected for IC/CMB from the large scale jet under the assumption that the X-ray emission is entirely due to IC/CMB. 

In Figure~3 we show the SED for each jet. The ALMA and \emph{Chandra} fluxes for the entire resolved jet outside the core, newly derived for this paper, are listed along with data from the literature in Table~1, where we also give the \emph{Fermi}/LAT (minimum) fluxes for the entire source. In Figure~3 the radio/sub-mm total jet fluxes are shown as black points with error bars, as are the \emph{Chandra} fluxes. The HST upper limits are also shown as black arrows. For the \emph{Fermi}/LAT observations we show three states of the source. In purple we show the `high state' SED made by compiling the 10 highest-TS bins from the original light-curve. In cyan we show the average SED as calculated from the entire time range of observations (10.5 years), and in dark yellow we give the minimum flux or upper limit for each of the \emph{Fermi}/LAT bands. Note that while the high and average state SEDs are generated from the same times on source, the low-state SED, by design, does not necessarily consist of fluxes measured from the same time-on-source. For OJ~287, the lowest flux/upper limit for each energy band occurred at combined bins 17, 4, 7, 10 and 16 from lowest to highest energy, respectively. For \pf, the values were taken from combined bins 7, 7, 1, 1, and 10.

In the IC/CMB scenario, the inverse Compton spectrum is essentially a copy of the synchrotron spectrum, shifted in frequency and luminosity according to the formulae in \cite{geo06} -- see also the discussion of the essential consistency between shifting the phenomenological curve and a more detailed physical model in \cite{meyer17}. As can be seen in Figure~3, the radio-optical synchrotron spectrum is reasonably well-constrained for both jets 
and the \emph{Fermi}/LAT minimum flux values agree extremely well with the IC/CMB model predictions based on the radio to X-ray data for both jets.

\section{Discussion and Conclusions}

It has been nearly two decades since the first papers suggesting that the anomalously high X-ray emission from the jet of PKS~0637$-$752 could be due to the IC/CMB mechanism were published. In the intervening decades several dozen quasar jets have had their high X-ray fluxes explained in this way, though it was generally impossible to rule out the alternative scenario of a second synchrotron component or hadronic models. In more recent years the IC/CMB explanation for the X-ray emission in most of these cases has been ruled out via deep gamma-ray upper limits \citep{meyer14,meyer15,meyer17,breiding2017,breiding2018_thesis}; the interpretation being that most kpc-scale jets are not as fast and/or not as aligned as generally required in the IC/CMB model.  

As originally explained in \cite{geo06}, the simplicity of the IC/CMB mechanism makes the predicted gamma-ray flux level inflexible given a well-sampled radio-optical synchrotron spectrum and a well-measured X-ray flux. The IC/CMB spectrum from the X-rays to gamma-rays has a shape set to match the radio-optical, and the normalization set by the X-rays. This normalization translates directly to a fixed value of B/$\delta$ where B is the magnetic field strength and $\delta$ the Doppler factor. 
Regardless of whether IC/CMB can explain the observed X-ray emission, it is a mandatory process (ambient photons will inevitably be upscattered to higher energies) and will produce X-ray to gamma-ray emission at some level.

\begin{figure}[t]
    \centering
    \includegraphics[width=3.5in]{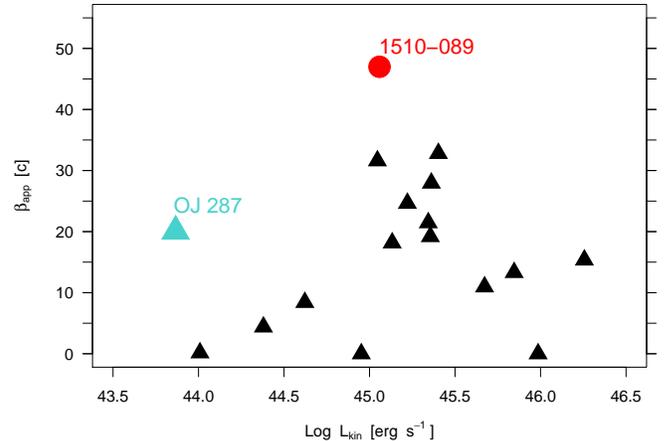}
    \caption{A comparison of OJ~287 and \pfs (cyan and red points) to 15 other X-ray jets for which IC/CMB was ruled out by \emph{Fermi}/LAT upper limits. On the x-axis is the jet kinetic power as scaled from low-frequency radio observations \citep[e.g.,][]{meyer11} and on the y-axis is the fastest recorded VLBI jet speed. Previous authors have noted an `envelope' in this plane with a forbidden zone at upper left \citep[e.g.][]{kharb2010}. This can be explained as a relation between $L_\mathrm{kin}$ and the intrinsic jet speed $\Gamma$, where sources at the upper edge of the envelope (which would run from lower left to upper right) are at the `critical angle' of $\theta\sim 1/\Gamma$. 
    }
    \label{fig:Bapp_Power}
\end{figure}
In this letter, we argue that we have detected the IC/CMB emission from OJ~287 and \pfs based on the excellent agreement between the predicted level of IC/CMB based on the radio through X-ray SED and the minimum flux level in the GeV band. The case is particularly strong for OJ~287 where we observe a plateau signature in the recombined light-curve which is consistent with a steady floor to the GeV flux.  The case for \pfs is less strong due to the lack of this signature -- it is possible that we have simply measured the minimum flux level of the core from the last ten years of \emph{Fermi/LAT} observations which coincidentally matches the level and spectrum expected from the jet under IC/CMB. Continued long-term monitoring by the \emph{Fermi}/LAT will clarify this issue. If the minimum-flux state currently measured is due to the core, then it is likely that in the next several years the core will reach even lower flux states as part of its overall variability. In that case we would see the measured GeV minimum-flux begin to dip down below the prediction shown in Figure~3, and we would have to rule out IC/CMB as the source of the X-ray emission in this source. On the other hand, if the minimum-flux state really is IC/CMB, then future observations of the source when the core is (inevitably) in a low state \emph{below} the steady emission of the large-scale jet will produce a plateau in the progressive binning curve. At present we favor the IC/CMB interpretation of the GeV minimum state in \pfs simply because it would be unlikely for the core to match the expected level and spectral shape by chance. It also seems likely that the flux measured in the 10-100 GeV band for \pf, which is somewhat above the IC/CMB prediction, is contaminated by the core (the minimum value in that band comes from the most time-on-source at 10 combined bins compared to only 1 bin in the previous two bands).

\begin{deluxetable*}{llllll}[th]
\tabletypesize{\small}
\tablecaption{\label{apptab1}Flux Measurements}
\centering
\tablehead{ Source & $\nu$ (Hz) & $\nu F_{\nu}$  & Source & $\nu$ (Hz) & $\nu F_{\nu}$  \\
 & (Hz) & (erg s$^{-1}$ cm$^{-2}$) &  & (Hz) & (erg s$^{-1}$ cm$^{-2}$) }
\startdata    
OJ~287  & 1.4 $\times$ 10$^{9}$   &  1.52 $\times$ 10$^{-16}$ (a)         & \pfs &  1.5 $\times$ 10$^{9}$  &  3.15 $\times$ 10$^{-15}$ \\
        & 4.8 $\times$ 10$^{9}$   &  1.98 $\times$ 10$^{-16}$ (a)        &            &  4.9 $\times$ 10$^{9}$  &  4.53 $\times$ 10$^{-15}$ \\
        & 1.5 $\times$ 10$^{10}$  &  2.14 $\times$ 10$^{-16}$ (a)         &            &  1.45 $\times$ 10$^{11}$  &  1.21 $\times$ 10$^{-14}$ \\
        & 5.2 $\times$ 10$^{14}$  &  $<$1.04 $\times$ 10$^{-16}$ (a)      &               &2.33 $\times$ 10$^{11}$  &  1.07 $\times$ 10$^{-14}$  \\
        & 2.4 $\times$ 10$^{17}$  &  2.49 $\pm$ 0.07 $\times$ 10$^{-14}$  &            &  3.43 $\times$ 10$^{11}$  &  1.00 $\times$ 10$^{-14}$  \\
        & 4.19 $\times$ 10$^{22}$  &  2.40 $\pm$ 0.78 $\times$ 10$^{-12}$&            &  5.1 $\times$ 10$^{14}$  &  $<$4.08$\times$ 10$^{-16}$ (b)  \\
        & 1.32 $\times$ 10$^{23}$  &  2.98 $\pm$ 1.09 $\times$ 10$^{-12}$ &            & 2.4 $\times$ 10$^{17}$  &  3.72 $\pm$ 0.05 $\times$ 10$^{-14}$   \\
        & 4.19 $\times$ 10$^{23}$  &  1.67 $\pm$ 0.68 $\times$ 10$^{-12}$&            & 4.19 $\times$ 10$^{22}$  &  2.36 $\pm$ 0.25 $\times$ 10$^{-11}$  \\
        & 1.32 $\times$ 10$^{24}$  &  1.38 $\pm$ 0.70 $\times$ 10$^{-12}$&            & 1.32 $\times$ 10$^{23}$  &  2.09 $\pm$ 0.19 $\times$ 10$^{-12}$  \\
        & 7.65 $\times$ 10$^{24}$  &  $<$6.52 $\times$ 10$^{-13}$      &            & 4.19 $\times$ 10$^{23}$  &  9.35 $\pm$ 4.31 $\times$ 10$^{-12}$ \\
        &                      &                               &         & 1.32 $\times$ 10$^{24}$  &  $<$4.69$\times$ 10$^{-12}$ \\
        &                      &                               &         & 7.65 $\times$ 10$^{24}$  &  2.12 $\pm$ 1.37 $\times$ 10$^{-12}$ \\
\enddata
\tablenotetext{a}{\cite{marscher2011}}
\tablenotetext{b}{\cite{sambruna2004}, otherwise this paper}
\end{deluxetable*}

The detection of IC/CMB implies a value for B/$\delta$ for each source. We have calculated equipartition values of B$\delta$ of 2.9 $\times$ 10$^{-5}$ G and 6.1 $\times$ 10$^{-5}$ G for OJ~287 and \pf, respectively, based on the 1.4 and 5 GHz radio observations and an assumed minimum electron Lorentz factor $\gamma_\mathrm{min}$ of 10. Using these values, the required Doppler factors $\delta$ are 22.5 and 19 to match the gamma-ray minimum level.

Given the results of our larger study of X-ray jets which are generally \emph{not} emitting X-rays dominated by IC/CMB, one must ask what makes these sources so different. While IC/CMB should increase significantly with redshift due to the $(1+z)^4$ enhancement of the CMB, these sources are not high-redshift at $z\sim0.3$. However, it is notable that each is highly superluminal on parsec scales, especially relative to their jet powers. OJ~287 has a maximum observed jet speed from Very Long Baseline Interferometry (VLBI) measurements of 20.1$c$ \citep{homan2001}, while \pfs has a maximum of 47$c$ \citep{jorstad2005}. Such high values immediately imply very small maximum orientation angles of 5.7$^\circ$ and 2.4$^\circ$ respectively. Further, when compared with other {Fermi}-detected X-ray jets, these speeds make them clear outliers. In Figure~4 we plot the apparent speed versus kinetic jet power for 15 jets where \emph{Fermi}/LAT has \emph{ruled out} IC/CMB as the source of the X-ray emission (data taken from \citealt{breiding2018_thesis} and forthcoming Breiding et al., 2019) as black triangles. The two subjects of this paper are noted as a cyan triangle and red circle, labeled.  

We argue that these sources are rare cases of orientation at the critical angle near 1/$\Gamma$ where $\beta_\mathrm{app}$ is maximized. In such cases the jet is just misaligned enough so that the large-scale jet is visible as an arcsecond-scale jet in high-resolution imaging, while being aligned enough that the (inevitable) IC/CMB emission is boosted significantly and dominates over the (presumed) synchrotron X-ray emission. The sensitivity to angle is considerable. As an illustration, let us assume that the jets have Lorentz factors of 22 and 50 and are oriented at the critical angle of 1/$\Gamma$, or 2.6$^\circ$ and 1.1$^\circ$ for OJ~287 and \pf, respectively. The beaming pattern for IC/CMB emission scales as $\delta^{4+2\alpha}$ (here we let $\alpha$=0.5), so at the assumed $\Gamma$ values the orientation angles of the two sources would need to increase to only 3.8$^\circ$ and 1.6$^\circ$, respectively for the IC/CMB X-ray emission to drop to 1/10th of the observed value. Given that the majority of similar (presumably more misaligned) sources in our larger sample appear to be dominated by synchrotron X-ray emission, this would likely dominate over the IC/CMB emission at most orientations. 

There is another factor which likely makes these jets outliers in addition to the orientation angle. As noted previously, population-based evidence as well as individual limits on $\delta$ from both proper motions \citep[in 3C~273;][]{meyer16} and deep \emph{Fermi}/LAT limits on the IC/CMB emission imply most large-scale jets are only mildly relativistic. Here we require high values of the Lorentz factor on extremely large scales -- assuming the maximum angle, the jets in OJ~287 and \pfs deproject to lengths of at least 600 and 800 kpc.  It is possible that only a few jets are able to accelerate flows and maintain high $\Gamma$ values at such distances. Such lengths also imply that these jets are not young. Interestingly, recent work on realistic GRMHD models of jets have trouble producing values of $\Gamma$ as high as are implied in the population (i.e., $\Gamma >$ a few), except where empty `channels' have been previously excavated, presumably by earlier jet activity (e.g., \citealt{bromberg2016}; see also the discussion in \citealt{marscher2011}). 

 In many cases the IC/CMB model requires super-Eddington jet power, but we find that is not required here, as was also found previously by \cite{marscher2011} for OJ~287. The minimum required power under the IC/CMB model is $1.8\times 10^{45}$ and $1.2\times 10^{45}$~erg~s$^{-1}$ for OJ~287 and \pfs respectively.
 Estimates of the black hole masses for OJ~287 and \pfs are moderately but not extremely large, at $10^{8.79}$ and $10^{8.62}$ $M_\odot$ \citep[][scaled from H$\beta$ width]{wang2004} yielding an Eddington luminosity of 5$-$8 $\times$ 10$^{46}$~erg~s$^{-1}$, which leaves a comfortable margin for the power requirements of the IC/CMB model even if one relaxes significantly away from the minimum. Interestingly, the estimated kinetic power of these jets, scaled from the low-frequency radio luminosity (\citealt{meyer11}, also Keenan et al., in prep.) is 7.4 $\times$ 10$^{43}$~erg~s$^{-1}$ and 1.1 $\times$ 10$^{45}$~erg~s$^{-1}$ for OJ~287 and \pf, respectively.  While the latter is a good match to the IC/CMB power requirements, for OJ~287 the discrepancy is considerable. However it has been shown that environment likely plays an important role in the large scatter of jet power/radio luminosity scalings \citep{hardcastle2013} and it is not clear that the low-frequency radio estimates of power are reliable for individual sources. 
 
Ultimately, it seems likely based on the full body of recent results which rule out the IC/CMB model for most X-ray jets, that OJ~287 and \pfs will prove to be outliers, with a particularly favorable alignment and unusually high jet speed at large distances. Indeed, it seems sources with Lorentz factors of 10-20 on nearly Mpc scales must be very rare, or else a similar case would have been confirmed much earlier, had the angle to the line of sight been even more favorable: halving the angle for either of these jets would produce a shorter, but still visible and extremely bright X-ray jet and even higher steady gamma-ray flux. Given the $(1+z)^4$ enhancement of the CMB and larger volume probed with redshift, it is likely that high-redshift counterparts to these unusual jets will eventually be found -- a promising candidate is the recently detected X-ray jet with very weak radio emission at $z$=2.5 \citep{sim2016}.

\acknowledgments
\noindent
E.T.M acknowledges NASA Fermi grants NNX15AU78G and 80NSSC18K1733.~E.T.M. and M.G. acknowledge NASA ADP grant NNX15AE55G and NSF grant 1714380.

\noindent
This paper makes use of the following ALMA data: 
\begin{itemize}
  \setlength{\parskip}{0pt}
  \setlength{\itemsep}{0pt plus 1pt}
    \item ADS/JAO.ALMA\#2016.1.00406.S
    \item ADS/JAO.ALMA\#2016.1.00116.S
\end{itemize}
\vspace{-\topsep}
\noindent
ALMA is a partnership of ESO (representing its member states), NSF (USA) and NINS (Japan), together with NRC (Canada), MOST and ASIAA (Taiwan), and KASI (Republic of Korea), in cooperation with the Republic of Chile. The Joint ALMA Observatory is operated by ESO, AUI/NRAO and NAOJ. The National Radio Astronomy Observatory is a facility of the National Science Foundation operated under cooperative agreement by Associated Universities, Inc.

%

\vspace{5mm}
\facilities{VLA, ALMA, HST, Chandra, Fermi}







\end{document}